\documentclass[twocolumn,a4paper,preprintnumbers,amsmath,amssymb,nofootinbib,floatfix, aps, showpacs]{revtex4}
\usepackage{pgfplots}
\usepackage[british]{babel}
\usepackage[utf8]{inputenc}
\usepackage{graphicx}% Include figure files
\usepackage{dcolumn}% Align table columns on decimal point
\usepackage{bm}% bold math
\usepackage{latexsym}
\usepackage{epsfig}
\usepackage{pbox}
\usepackage{multirow}
\usepackage{rotating}

\usepackage{textcomp}

\usepackage{color}
\usepackage{mathrsfs}
\usepackage{bbm}
\usepackage{bm}% bold math
\usepackage{amsmath,amsfonts,amssymb,textcomp}
\usepackage{mathrsfs}
\usepackage{placeins}
\usepackage{xcolor}
\usepackage{hyperref}
\hypersetup{colorlinks=true,linkbordercolor=blue,linkcolor=blue, citecolor=blue}
\usepackage{epsfig}

\usepackage{bbold}
\usepackage{slashed}
\usepackage{ifpdf}

\newcommand{\be}{\begin{equation}}
\newcommand{\ee}{\end{equation}}

\begin{document}

\title{Mass dimension one fermions and their gravitational interaction}

\author{R. J. Bueno Rogerio$^{1}$}\email{rodolforogerio@unifei.edu.br}
\author{R. de C. Lima$^{2}$}\email{rodrigo.lima@feg.unesp.br}
\author{L. Duarte$^{3}$}\email{Laura.Duarte@ific.uv.es}
\author{J. M. Hoff da Silva$^{4}$}\email{julio.hoff@unesp.br}

\affiliation{$^{1}$Instituto de F\'isica e Qu\'imica, Universidade Federal de Itajub\'a - IFQ (UNIFEI), \\
Av. BPS 1303, CEP 37500-903, Itajub\'a - MG, Brazil.}
\affiliation{$^{2,3,4}$Departamento de F\'isica e Qu\'imica, Universidade Estadual Paulista (UNESP),\\Guaratinguet\'a-SP-Brazil}
\affiliation{$^{3}$AHEP Group, Institut de F\'isica Corpuscular (CSIC-Universitat de Val\`encia), Parc Cientific de Paterna.
C/Catedratico Jos\'e Beltr\'an, 2 E-46980 Paterna (Val\`encia) - Spain}

\author{M. Dias$^{5}$}\email{marco.dias@unifesp.br}
\author{C. R. Senise Jr.$^{6}$}\email{carlos.senise@unifesp.br}
\affiliation{$^{5,6}$Departamento de F\'isica, Universidade Federal de S\~ao Paulo (UNIFESP),\\Diadema-SP-Brazil}

%\maketitle
%\keywords{}
%\bigskip

\begin{abstract}
We investigate in detail the interaction between the spin-${1/2}$ fields endowed with mass dimension one and the graviton. We obtain an interaction vertex that combines the characteristics of scalar-graviton and Dirac's fermion-graviton vertices, due to the scalar-dynamic attribute and the fermionic structure of this field. It is shown that the vertex obtained obeys the Ward-Takahashi identity, ensuring the gauge invariance for this interaction. In the contribution of the mass dimension one fermion to the graviton propagator at one-loop, we found the conditions for the cancellation of the tadpole term by a cosmological counter-term. We calculate the scattering process for arbitrary momentum. For low energies, the result reveals that only the scalar sector present in the vertex contributes to the gravitational potential. Finally, we evaluate the non relativistic limit of the gravitational interaction and obtain an attractive Newtonian potential, as required for a dark matter candidate.
\end{abstract}

\maketitle

\section{Introduction}

The lack of theoretical basis to approach the dark matter problem is usually inputed to several, and somewhat concomitant, reasons. In fact, the complexity and peculiarity of such a problem has taken the scientific community to a plethora of attempts in the investigation of this phenomena. These attempts range from extra dimensions to more conservative geometric setups, from phenomenological cosmology to modeling based in exotic potentials. While the problem remains unsolved, an approach based on solid criteria erected from quantum field theory is particularly sound. This paper intend to study some relevant aspects of the interaction between such a candidate and weak field gravity.  

Proposed in its first version in 2004 \cite{Ahluwalia:2004ab}, the fermionic field of spin-${1/2}$ with mass dimension one is constructed upon a complete set of eigenspinors of the charge conjugation operator, the Elkos. In the early formulation, these fields were quantum objects carrying a representation of subgroups of the Lorentz group $H\!O\!M\!(2)$ and $S\!I\!M\!(2)$ \cite{CG}, and corresponding semi-direct extension encompassing translation. Quite recently, however, a modification in the spinor dual - taking advantage of the fact that in spinor physics only a bilinear is observable - has lead to a theory endowed with full Lorentz (Poincar\`e) symmetry. The main steps of this formulation, along with the theory of duals may be found\footnote{Occasionally, as the situation requires, we shall evince one or another necessary point of the dual formulation.} in Ref. \cite{ahluwa1}. These formulations boosted several works in a broad range of areas, for instance mathematical physics \cite{daRocha:2007pz,daRocha:2011yr,fab,fab2,daRocha:2011xb,daRocha:2013qhu,HoffdaSilva:2016ffx}, phenomenology of particles \cite{Alves:2014kta,Agarwal:2014oaa,Alves:2014qua,CYL,CYL2,Alves:2017joy} and cosmology \cite{daSilva:2014kfa,arra,urru,oqui,sac,eh,nos,so,cheg,Pereira:2014wta,Pereira:2014pqa,Pereira:2017efk,Pereira:2016eez,Rogerio:2017gvr}. 

Let us to pinpoint the main physical aspects of this recent formulation and its insertion in the irreducible representation of the Poincar\`e symmetries.        

The prominent work due to Wigner, scrutinizing the physical content supported by the Hilbert space under the Poincar\`e group action \cite{Wig1}, found consistently one particle states. Nevertheless all the investigation was performed within the proper orthochronous Lorentz subgroup. In a less known work, Wigner generalized the investigation to the inhomogeneous Lorentz group as a whole, by including discrete symmetries \cite{Wig2}. As a result, hidden particle classes appear and it turns out that the particle studied in \cite{ahluwa1} behaves, under discrete symmetries, in a way predicted in one of these cases. Having said that, we shall now depict the precise construction whose consideration enables the field to undergo a dynamics that leads to the mass dimension one property.          

The construction of the field as a spin-${1/2}$ representation is characterized, of course, by the presence of spinors as expansion coefficients. These spinors, as usual, belongs to the $(0,1/2)\oplus (1/2,0)$ Weyl representation space. This is indeed the prescription for Dirac spinors. The crucial difference arises in the way that both sectors of the representation space are related. For Dirac spinors parity is used, and as an inexorable and direct consequence the Dirac dynamics is reached. This relation is literal: in acting on spinors, the parity operator is the Dirac operator \cite{LE}, and vice-versa \cite{EL}. As a consequence, being the different sectors of the representation space related by means of another procedure (where parity plays no role), the Dirac dynamics is no longer expected. Since the construction is relativistic, one is forced to conclude that the Klein-Gordon dynamics is in fact in order. Finally, the quantum field shall inherit such a dynamics, from which the canonical mass dimension is read. The combined characteristics of mass dimension one, along with eigenspinors of the charge conjugate operator perform the neutrality of the field.     

In this work we study the interaction between this fermionic field and the graviton, by means of the weak field approximation at first order in the background expansion parameter, from the covariant action of gravity. The paper is organized as follows: in the next section, starting from the mass dimension one spin-${1/2}$ 
fermionic field action in curved spacetime, we obtain the first two interaction vertex from which we evince the Ward-Takahashi identity. Taking advantage of the results of the previous section, with the one graviton vertex at hands, we compute the one-loop correction for the graviton propagator and study the tadpole counter-term responsible to remove the divergent part of the interaction. Finally, we delve into the study of the gravitational scattering process in the non-relativistic limit, obtaining the attractive Newtonian potential, as required for a dark matter candidate. 

\section{Mass dimension one fermion-graviton interaction vertex}\label{MDOFGIV}

The action for the mass dimension one spin-${1/2}$ fermionic field in a curved background can be written\footnote{The notation for the quantum field used throughout this paper shall not be confused with the expansion coefficients of the quantum field in Ref. \cite{ahluwa1}.} as \cite{daSilva:2014kfa,Pereira:2016eez}
\begin{equation}\label{v1}
 \mathcal{S} = \int\sqrt{-g}\left(g^{\mu\nu}\nabla_{\mu}\tilde{\lambda}\nabla_{\nu}\lambda - m^2\tilde{\lambda}\lambda\right)d^{4}x,
\end{equation}
where $\lambda = \lambda(x)$ and $\tilde{\lambda} = \tilde{\lambda}(x)$ represent the spinor field and its corresponding dual, coupled to gravity. The metric determinant is denoted as usual by ${g \equiv det (g_{\mu\nu})}$, with metric signature ${(+,-,-,-)}$. The covariant derivatives act in the 
fermionic fields as
\begin{equation}\label{x2}
\nabla_{\mu}\tilde{\lambda} = \partial_{\mu}\tilde{\lambda} + \tilde{\lambda}\Gamma_{\mu} \textnormal{ and }  \nabla_{\mu}\lambda = \partial_{\mu}\lambda - \Gamma_{\mu}\lambda,
\end{equation}
with the spin connection defined as ${\Gamma_{\mu} = A_{\mu ab}\sigma^{ab}}$. Also, the generators of transformations, ${\sigma^{ab}=-1/2[\gamma^{a},\gamma^{b}]}$, are written in terms of a tetrad field\footnote{Constructed usually as ${g_{\mu\nu}(x) = e_{\mu}^{a}(x)e_{\nu}^{b}(x)\eta_{ab}}$, such as ${e^{\mu}_{a}(x)e_{\nu}^{a}(x) = \delta_{\nu}^{\mu}}$ and ${e^{a}_{\mu}(x)e_{b}^{\mu}(x) = \delta_{b}^{a}}$.} ${e_{\alpha}^{a}}$ \cite{yepez} and the gamma matrices in the locally flat space. Finally, the term ${A_{\mu ab}}$ is given by
\begin{equation}\label{x1}
A_{\mu \hspace{0.75mm} b}^{\hspace{0.6mm}a}=-e_{b}^{\nu}\partial_{\mu}e_{\nu}^{a}+{e^{\nu}_{b}\Gamma_{\mu\nu}^{\alpha}e_{\alpha}^{a}}.
\end{equation}

Since the expansion in terms of the tetrad fields \(e^a_\mu\) can be connected with the same weak field expansion for the  metric around Minkowski space, we proceed with %(see appendix \ref{AA} for further details)
\begin{eqnarray}
\label{v3}&& g_{\mu\nu}= \eta_{\mu\nu}+\kappa h_{\mu\nu}, \quad |\kappa h_{\mu\nu}|\ll 1, \\
%\label{v33}&& \Gamma_{\mu}=A_{\mu ab}\sigma^{ab} \simeq =\left(e_{b}^{\rho}\Gamma^{\chi}_{\rho \alpha}e_{a \chi}-e_{b}^{\rho}\partial_{\alpha}{e_{a \rho}}\right) \sigma^{a b}/4,\\
\label{v302}&& e_{a}^{\alpha}=\eta_{a}^{\alpha}-1/2\kappa h_{a}^{\alpha}+3/8\kappa^2 h_{a}^{\chi }h_{\chi}^{\alpha}+ \mathcal{O}(\kappa^3),\\
\label{v303}&& e^{a}_{ \alpha} =\eta^{a}_{\alpha}+1/2\kappa h^{a}_{\alpha}-1/8\kappa^2h_{\alpha \chi} h^{a \chi}+ \mathcal{O}(\kappa^3),\\
\label{v31}&& g^{\alpha \beta} =\eta^{\alpha \beta}-\kappa h^{\alpha \beta}+\kappa^2 h^{\alpha \chi}h_{\chi}^{\beta}+ \mathcal{O}(\kappa^3),
\end{eqnarray}
where ${\kappa^{2} = 16\pi G}$ and the gamma matrices ${\gamma^{\alpha}}$ are written in the Weyl representation. Moreover, it is possible to write
\begin{eqnarray*}
\Gamma_\mu\!\! &=&\frac{\sigma^{\alpha\beta}}{4}\left[\kappa\partial_\beta h_{\mu\alpha}-\kappa\partial_{\alpha}h_{\mu\beta}+\frac{\kappa^2}{4}h^\rho_\beta\partial_\mu h_{\alpha\rho}\right.\nonumber
\\
&-&\left.\frac{\kappa^2}{4}h^\rho_\alpha\partial_\mu h_{\beta\rho} +\frac{\kappa^2}{4}h^\rho_\beta\partial_\alpha h_{\mu\rho}-\frac{\kappa^2}{4}h^\rho_\alpha\partial_\beta h_{\mu\rho}\right.
\\\nonumber
&+&\left.\frac{\kappa^2}{4}h^\rho_\alpha\partial_\rho h_{\mu\beta}-\frac{\kappa^2}{4}h^\rho_\beta\partial_\rho h_{\mu\alpha}\right].\nonumber
\end{eqnarray*}

Rewriting the lagrangian density read from Eq. (\ref{v1}), neglecting terms at orders higher than ${\kappa^3}$ and using the previous relations in the momentum space representation, the one and two-interaction vertices are obtained by performing the functional variation. By doing so, one has the mass dimension one fermion and one graviton interaction vertex in the tree level described by
\begin{eqnarray}\label{14}
V_{\alpha\beta}(p,q,r)&=&i\frac{\kappa}{8}\delta(q-r-p)\Big[4(p\cdot q - m^{2})\mathbbm{1}\eta_{\alpha\beta} 
\nonumber\\
&-& 4(q_{\alpha}p_{\beta}+q_{\beta}p_{\alpha})\mathbbm{1} 
+ [\gamma_{\alpha},\gamma_{\mu}r^{\mu}](p+q)_{\beta}\nonumber\\
&+&[\gamma_{\beta},\gamma_{\nu}r^{\nu}](p+q)_{\alpha}\Big].
\end{eqnarray}
In the expression above, $\mathbbm{1}$ stands for the identity matrix, $p$ is the incoming momentum related to $\lambda$, $q$ is the  outcome momentum associated with $\tilde{\lambda}$ and $r$ is the incoming momentum for the graviton field. Notice that the vertex (\ref{14}) is composed by two parts, the first one, [$4(p\cdot q - m^{2})\mathbbm{1}\eta_{\alpha\beta} - 4(q_{\alpha}p_{\beta}+q_{\beta}p_{\alpha})\mathbbm{1}$], typically performing a scalar-graviton sector and the second one, whose terms are typical of a fermion-graviton interaction \cite{holstein}. This peculiar behavior is also verified in other contexts when studying mass dimension one spinors \cite{sauloeu,saulao}. The reason for that rests upon the combined character of a dynamic governed by the Klein-Gordon equation, while having simultaneously a spinorial structure for the field. 

In order to assert gauge invariance for the interaction between mass dimension one fermions and the graviton we shall finalize this section by pointing out the Ward-Takahashi \cite{ward,Capper:1974vb} relation for this specific case
\begin{eqnarray}
2r^{\beta}V_{\alpha\beta}(p,q) = W_{\alpha}(p,q),  
\end{eqnarray}
where $W_{\alpha}(p,q) = p_{\alpha}S(q)-q_{\alpha}S(p)+r^{\beta}\{\omega_{\alpha\beta}S(q)
+S(p)\omega_{\alpha\beta}\}$, with $\omega_{\alpha\beta}=-\omega_{\beta\alpha}=\frac{1}{4}\sigma_{\alpha\beta}$ stands for the transformation parameter, $r\equiv p-q$, and 
\begin{eqnarray}
S(p) = i(p^2-m^2)\mathbbm{1}
\end{eqnarray}
is related to the mass dimension one fermion propagator given in \cite{ahluwa1}. Using the vertex presented in Eq. (\ref{14}) it is possible to find the correct numerical factor in the Ward-Takahashi relation, necessary to the case at hand. 

\section{One-loop mass dimension one fermion divergences on the graviton propagator}\label{OLMOFCGP}

\begin{figure}[!ht]
\centering
\begin{minipage}{0.29\textwidth}
   \includegraphics[width=\textwidth]{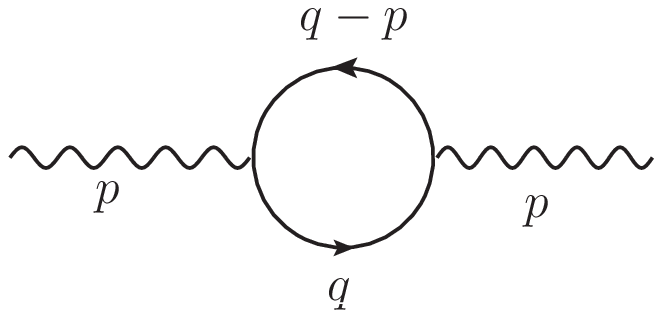}
\end{minipage}
\begin{minipage}{0.13\textwidth}
    \includegraphics[width=\textwidth]{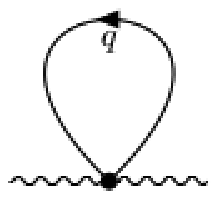}
\end{minipage}

\caption{Two graphs, (a)-left and (b)-right for
mass dimension one contribution to the graviton self-energy. }
\label{cm1}
\end{figure}

Using the vertex (\ref{14}), the two graviton vertex, and the mass dimension one propagator it is possible to evaluate the graphs outlined in Fig. (\ref{cm1}), both involving one-loop contributions to the graviton self-energy. Throughout this section use is made of dimensional regularization with $d=4-2\epsilon$. 

In terms of the Barnes-Rivers operators \cite{accioly}, the divergent part of the one loop correction for graviton self-energy 
contribution (without a tadpole term) due to the graph (a) of Fig.  (\ref{cm1}) is given by 
\begin{eqnarray}
\Pi^{\mu\nu,\alpha\beta}_{(a)}=-\int\frac{d^d q}{(2\pi)^d}\frac{Tr\left[V^{\mu\nu}i\mathbbm{1}V^{\alpha\beta}i\mathbbm{1}\right]}{(q^2-m^2)[(p-q)^2-m^2]},\label{ra}
\end{eqnarray} where $V^{\mu\nu} = V^{\mu\nu}(q-p,q,p)$ and $V^{\alpha\beta} = V^{\alpha\beta}(q,q-p,-p)$, resulting in 
\begin{eqnarray}
\Pi^{\mu\nu,\alpha\beta}_{(a)}&=&\frac{1}{\pi^2\epsilon}\left(\frac{m^4}{2\cdot 16}-\frac{8m^2p^2}{3\cdot 16}\right){P^0}^{\mu\nu,\alpha\beta}\nonumber\\
&+& \frac{1}{\pi^2\epsilon}\left(-\frac{m^4}{ 16}-\frac{m^2p^2}{ 16} \right) {P^1}^{\mu\nu,\alpha\beta}\nonumber\\
&+& \frac{1}{\pi^2\epsilon}\left(-\frac{5m^2p^2}{3\cdot 8}-\frac{m^4}{ 16} +\frac{3p^4}{10\cdot 16}\right){P^2}^{\mu\nu,\alpha\beta}\nonumber\\
&+&\frac{1}{\pi^2\epsilon}\left(-\frac{m^4}{2\cdot 16} \right) {\bar{P^0}}^{\mu\nu,\alpha\beta}\nonumber\\
&+&\left(\frac{m^4}{2\cdot 16}\right){\bar{\bar{P^0}}}^{\mu\nu,\alpha\beta},\label{evok0}
\end{eqnarray}
%where the divergent part of some integrals used here are listed in Appendix \ref{AAAA} (for readily reference), 
computed through the Feynman parametrization technique
${(ab)^{-1} = \int_{0}^{1} [az + b(1-z)]^{-2}dz}$ \cite{Ryder} applied in the context of weak field gravity \cite{capper, DeMeyer:1974ed}. Similarly, the contribution coming from the graph (b) reads 
\begin{eqnarray}
\Pi^{\mu\nu,\alpha\beta}_{(b)}&=&-\int\frac{d^d q}{(2\pi)^d}\frac{Tr\left[V^{\mu\nu\alpha\beta}i\mathbbm{1}\right]}{(q^2-m^2)},\label{rara}
\end{eqnarray} 
where $V^{\mu\nu\alpha\beta} = V^{\mu\nu\alpha\beta}(r,-s,q,-q)$. This last contribution may be computed by means of the two graviton vertex ${V^{\mu\nu\alpha\beta}(r,s,p,q)}$. Here we depict directly the total contribution 
\begin{eqnarray}
\Pi^{\mu\nu,\alpha\beta}_{(a)+(b)}&=&-\frac{1}{16\pi^2\epsilon}\left(m^4-2m^2p^2\right){P^0}^{\mu\nu,\alpha\beta}\nonumber\\
&+&\frac{1}{16\pi^2\epsilon}\left(2m^4+m^2p^2 \right) {P^1}^{\mu\nu,\alpha\beta}\nonumber\\
&+&\frac{1}{16\pi^2\epsilon}\left( 2m^4+2m^2p^2\right){P^2}^{\mu\nu,\alpha\beta}\nonumber\\
&+&\frac{1}{16\pi^2\epsilon}\left(m^4\right) {\bar{P^0}}^{\mu\nu,\alpha\beta}\nonumber\\
&-&\frac{1}{16\pi^2\epsilon}\left(m^4\right){\bar{\bar{P^0}}}^{\mu\nu,\alpha\beta}.\label{evok1}
\end{eqnarray}

\begin{figure}[!ht]
\centering
\includegraphics[scale=0.6]{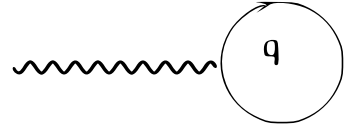}
\caption{Tadpole graph.}
\label{fig2}
\end{figure}

Now we shall focus on the tadpole term (and corresponding contribution), whose graph is shown in Fig. (\ref{fig2}). Its divergent part is simply given by 
\begin{eqnarray}
W_{\mu\nu}&=& -\kappa\frac{m^4}{16\pi^2\epsilon}\eta_{\mu\nu},\label{ram0}
\end{eqnarray} where we have used 
\begin{eqnarray*}
\int\frac{d^dq}{q^2-m^2}&=&\frac{i\pi^2m^2}{\epsilon}+\textnormal{finite terms},\nonumber\\
\int\frac{d^dq\, q^\mu q^\nu}{q^2-m^2}&=&\frac{i\pi^2 m^4}{4\epsilon}\eta^{\mu\nu}+\textnormal{finite terms}.
\end{eqnarray*}

It is well know that the functional generator $Z[g_{\alpha\beta}]=\int d[h_{\alpha\beta}]e^{i\int d^{4}x\mathcal{L}}$ is invariant under an usual changing of variables \({x^\prime}^\mu=x^\mu+\varepsilon^\mu \). Such an invariance leads to 
\begin{eqnarray}
\delta Z=\int d^4x \frac{\delta Z}{\delta g_{\alpha\beta}}\delta g_{\alpha\beta}=\int d^4x \hat{A}_{\alpha\beta\lambda}\frac{\delta Z}{\delta h_{\alpha\beta}}=0\label{rarara}
\end{eqnarray} and, by means of a simple integration by parts, we are lead to 
\begin{eqnarray*}
\hat{A}_{\alpha\beta\lambda}=-\eta_{\alpha\lambda}\partial_\beta-\kappa(h_{\alpha\lambda}\partial_\beta+\partial_\beta h_{\alpha\lambda}-\frac{1}{2}\partial_\lambda h_{\alpha\beta}).%\label{kiki}
\end{eqnarray*} 
Taking the functional derivative of $Z[g_{\alpha\beta}]$ with respect to $h_{\rho\sigma}$ and converting the result to the momentum space, after using the relations   
\begin{eqnarray}\label{ragnarock2}
&&p^\mu P^2_{\mu\nu,\rho\sigma}=p^\mu P^0_{\mu\nu,\rho\sigma}=0,\nonumber\\
&&p^\mu P^1_{\mu\nu,\rho\sigma}=\frac{1}{2}\left(\Theta_{\nu\rho}p_\sigma+\Theta_{\nu\sigma}p_\rho\right),\nonumber\\
&&p^\mu \bar{\bar{P^0}}_{\mu\nu,\rho\sigma}=p_\nu\Theta_{\rho\sigma},\nonumber\\
&&p^\mu \bar{P^0}_{\mu\nu,\rho\sigma}=p_\nu\omega_{\rho\sigma},
\end{eqnarray} 
 we get the Ward identity:
\begin{eqnarray}
p_\mu\Pi^{\mu\nu,\rho\sigma}+\frac{\kappa}{2}\left(\eta^{\nu\rho}p_\mu W^{\sigma\mu}+\eta^{\nu\sigma}p_\mu W^{\rho \mu}-p^\nu W^{\rho\sigma}\right)=0.\nonumber\\* \label{bouba}
\end{eqnarray} As an important consistency check we remark that Eq. (\ref{bouba}) is indeed satisfied for $\Pi^{\mu\nu,\rho\sigma}=\Pi_{(a)}^{\mu\nu,\rho\sigma}+\Pi_{(b)}^{\mu\nu,\rho\sigma}$ along with the tadpole term contribution presented in Eq. (\ref{ram0}), as expected. 

We remark that, in fact, one can add a cosmological term to the original lagragian as done in Ref. \cite{capper}, 
\begin{equation}\label{refx1}
\mathcal{L'}=\mathcal{L}-\Lambda\sqrt{g},\end{equation}
in order to cancel out the tadpole contributions. The lagrangian (\ref{refx1}), in the first order formalism, lead to the recognition of $\Lambda$ as  $\Big[\Lambda=\kappa\frac{m^4}{8\pi^2\epsilon}\Big]$ in order to cancel the tadpoles contributions to the graviton self-energy via a new set of Feynman graphs introduced. 

We would like to finalize this Section with two parenthetically remarks. Certainly, in higher orders of \(\kappa\), other terms appear, deeply modifying the graviton self-energy, but we shall not explore these terms in this work. Also, as it is clear, we have a theory with more momentum terms in the propagator than in the Dirac fermionic case. This fact could lead to some concern about unitarity, in the sense that the theory at hand could be faced, then, as a 'higher dimensional' fermionic theory. We remark, however, that the mass dimension one fermions are also constrained by a subsidiary first order equation \cite{ahluwa1}, not dynamical at the present case, which could help providing the right factorization ensuring unitarity.        

\section{Gravitational Potential in the non-relativistic limit}\label{GPNRL}

The computation of a scattering process can be accomplished by associating the respective Feynman diagrams in a certain order of perturbation. Such a procedure, as it is well known, reproduces the results of %non-
Relativistic Quantum
Mechanics, where the interaction between the particles is described by a potential $V(\bf{x})$. Since our interest here lies in the study of the physical content of the mass dimension one fermions and graviton interaction, we shall study in this section the non-relativistic limit of such an interaction, after performing the calculations of the full relativistic scattering amplitude of two mass dimension one fermions mediated by a graviton at tree level. As asserted before, the result provides an optimistic scenario in putting these spinors as candidates to (at last part of) dark matter, since the attractive Newtonian gravitational potential is reached. 

\begin{figure}[!ht]
\centering
\includegraphics[scale=0.6]{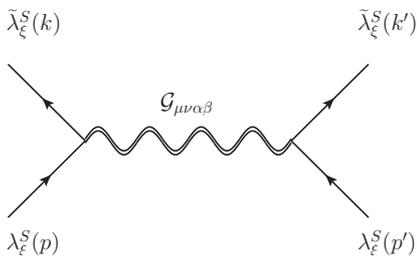}
\caption{Mass dimension one fermions scattering mediated by a graviton.}
\label{fig3}
\end{figure}

The relation between the potential $V({\bf x})$ and the scattering amplitude $\mathcal{M}$ is given by
\begin{eqnarray}\label{v}
 V({\bf x}) = \frac{-i}{2E_{1}}\frac{1}{2E_{2}}\int \frac{d^3 {\bf r}}{(2\pi)^3}\mathcal{M}({\bf r})e^{i{\bf r}\cdot{\bf x}},
\end{eqnarray}
where ${\bf r}$ stands for the exchanged graviton momentum. The Feynman graph associated with the scattering $\lambda^{S}_{\xi}({ p}) \lambda^{S}_{\xi}({p'}) \rightarrow \widetilde{\lambda}^{S}_{\xi}( { k})  \widetilde{\lambda}^{S}_{\xi}( {k'})$, mediated by a graviton, can be seen in Fig. (\ref{fig3}).
%\FloatBarrier 
The corresponding scattering amplitude is
\begin{eqnarray}\label{a}
\mathcal{M}=\!\! \frac{1}{m^2}\left(\widetilde{\lambda}^{S}_{\xi}(k) V^{\alpha\beta}\lambda^{S}_{\xi}({ p}) \mathcal{G}_{\mu\nu\alpha\beta}\widetilde{\lambda}^{S}_{\xi}(k') V^{\mu\nu}\lambda^{S}_{\xi}({ p'})\right),
\end{eqnarray}
where the quantities composing the amplitude are given by
\begin{eqnarray}
&&\widetilde{\lambda}^{S}_{\xi}(k)=\stackrel{\neg}{\lambda}^{S}_{\xi}(k)\mathcal{A}\label{duala}, \quad 
\widetilde{\lambda}^{A}_{\xi}(k)=\stackrel{\neg}{\lambda}^{A}_{\xi}(k)\mathcal{B},\\
&&\mathcal{G}_{\alpha\beta\mu\nu}=\frac{1}{2r^2}(\eta_{\alpha\mu}\eta_{\beta\nu}+\eta_{\beta\mu}\eta_{\alpha\nu}-\eta_{\alpha\beta}\eta_{\mu\nu}).
\end{eqnarray}%%%%%%%%%%%%
The term ${\mathcal{G}_{\alpha\beta\mu\nu}}$ is the graviton propagator in the momentum space, presented in Ref. \cite{capper}. Its construction and gauge fixing are presented in Ref. \cite{Capper:1973pv}. The lower-index ${\xi}$ appearing in (\ref{duala}) stands for the spinors helicity states $(\pm,\mp)$. The operators $\mathcal{A}$ and $\mathcal{B}$ are written\footnote{The operators $\mathcal{A}$ and $\mathcal{B}$ are a redefinition of the spinor dual, necessary for the theory to be Lorentz invariant. This can be verified by analyzing the spin sums in the limit ${\tau\rightarrow 1}$. Besides that, the orthonormality relations remain intact.} according to Ref. \cite{ahluwa1}
%\begin{eqnarray}
%\mathcal{A}=2\bigg(\frac{\mathbbm{1}-\tau G}{1-\tau^2}\bigg),\quad
%\mathcal{B}=2\bigg(\frac{\mathbbm{1}+\tau G}{1-\tau^2}\bigg),
%\end{eqnarray} 
%where $G$ is a matrix appearing (with well posed representation in the momentum space) in the new dual definition. Its explicit form won't be necessary here, only some of its properties. For a complete account on this object see \cite{ahluwa1}. 
We remark that in the definition of the $\mathcal{A}$ and $\mathcal{B}$ operators the limit $\tau\rightarrow 1$ is implicit. This parameter, as well as its limit, as we shall see, plays no role in our analysis. 

Both operators are governed by the following properties,
\begin{eqnarray}
\label{abop}\mathcal{A}\widetilde{\lambda}^{S}_{\xi}(k)=\widetilde{\lambda}^{S}_{\xi}(k),\quad\mathcal{B}\widetilde{\lambda}^{A}_{\xi}(k)=\widetilde{\lambda}^{A}_{\xi}(k),
\end{eqnarray}
needed to calculate the invariant amplitude. Analogously to the vertex given in (\ref{14}), we reach the following interaction vertex
\begin{eqnarray}\label{38}
V^{\mu\nu}&=&\frac{\kappa}{16}[\underbrace{4(k'\cdot p' - m^{2})\eta^{\mu\nu} - 4(k'^{\mu}p'^{\nu}+k'^{\nu}p'^{\mu})}_{E^{\mu\nu}}\nonumber\\
&+& \underbrace{[\gamma^{\mu},\slashed{r}](p'+k')^{\nu}+[\gamma^{\nu},\slashed{r}](p'+k')^{\mu}}_{M^{\mu\nu}}]\nonumber \\ 
          &=& \frac{\kappa}{16}(E^{\mu\nu}+M^{\mu\nu}).
\end{eqnarray}
In Eq. (\ref{38}), $E^{\mu\nu}$ and $M^{\mu\nu}$ stand for the scalar (with an implicit identity matrix) and fermionic sectors, respectively, composing the vertex. By inserting (\ref{38}) into (\ref{a}), 
%we are able to write
%\begin{align}
%\mathcal{M}= \frac{\kappa^2}{256m^2}\Big[\Big(\stackrel{\neg}{\lambda}^{S}_{\xi}(k)\mathcal{A} E^{\alpha\beta}\lambda^{S}_{\xi}({p})+\stackrel{\neg}{\lambda}^{S}_{\xi}(k)\mathcal{A} M^{\alpha\beta}\lambda^{S}_{\xi}({p})\Big)\nonumber \\
%\times\mathcal{G}_{\mu\nu\alpha\beta}\Big(\stackrel{\neg}{\lambda}^{S}_{\xi}(k')\mathcal{A} E^{\mu\nu}\lambda^{S}_{\xi}({p'})+\stackrel{\neg}{\lambda}^{S}_{\xi}(k')\mathcal{A} M^{\mu\nu}\lambda^{S}_{\xi}({p'})\Big)\Big].
 %\label{1}
%\end{align}
using the relations (\ref{abop}), the explicit form of the operator $\mathcal{A}$, knowing a important to highlight that using the identities $GM^{\mu\nu}=M^{\mu\nu}G$ and 
$G \lambda^{S}_{\xi}({p}) = \lambda^{S}_{\xi}({p})$, we 
obtain
\begin{eqnarray}\label{5}
&&\mathcal{M}=\frac{\kappa^2}{256m^2}\Big[\Big(E^{\alpha\beta}\stackrel{\neg}{\lambda}^{S}_{\xi}(k) \lambda^{S}_{\xi}({p})\nonumber\\
&&+\frac{2-2\tau}{1-\tau^2}\stackrel{\neg}{\lambda}^{S}_{\xi}(k) M^{\alpha\beta}\lambda^{S}_{\xi}({p})\Big)  \mathcal{G}_{\mu\nu\alpha\beta}\nonumber\\
&&\times\Big(E^{\mu\nu}\stackrel{\neg}{\lambda}^{S}_{\xi}(k') \lambda^{S}_{\xi}({p'})\nonumber\\
&&+\frac{2-2\tau}{1-\tau^2}\stackrel{\neg}{\lambda}^{S}_{\xi}(k') M^{\mu\nu}\lambda^{S}_{\xi}({p'})\Big)\Big].
\end{eqnarray} 
\begin{widetext}
It can be readily verified now that the aforementioned implicit limit $\tau \rightarrow 1$ is trivial. After contracting the propagator with one of the vertices, substituting $E^{\alpha\beta}$ and $M^{\alpha\beta}$, besides using the the standard mass dimension one spinors  ${(\lambda^{S}_{\{+,-\}},\stackrel{\neg}{\lambda}^{S}_{\{+,-\}})}$, we have
%\begin{strip}
\begin{eqnarray}\label{1111} 
&&\mathcal{M}=\mathcal{C}\Big[ 64m^2 (4m^2(p'\cdot k')-4m^4+2(k\cdot k')(p\cdot p'))+16m^3  \stackrel{\neg}{\lambda}^{S}_{\{+,-\}}(k)[\slashed{k'}+\slashed{p'}, \slashed{r}]\lambda^{S}_{\{+,-\}}(p')\nonumber\\
&&-16m k\cdot (k'+p')\stackrel{\neg}{\lambda}^{S}_{\{+,-\}}(k')\times[\slashed{p}, \slashed{r}]\lambda^{S}_{\{+,-\}}(p')-16m p\cdot (k'+p')\stackrel{\neg}{\lambda}^{S}_{\{+,-\}}(k')[\slashed{k}, \slashed{r}]\lambda^{S}_{\{+,-\}}(p')\nonumber\\
&&+16m[m^2\stackrel{\neg}{\lambda}^{S}_{\{+,-\}}(k)[\slashed{k'}+\slashed{p'}, \slashed{r}]\lambda^{S}_{\{+,-\}}(p)-p'\cdot (k+p)\times\stackrel{\neg}{\lambda}^{S}_{\{+,-\}}(k)[\slashed{k'}, \slashed{r}]\lambda^{S}_{\{+,-\}}(p)-k'\cdot (k+p)\stackrel{\neg}{\lambda}^{S}_{\{+,-\}}(k)\nonumber\\
&&\times[\slashed{p'}, \slashed{r}]\lambda^{S}_{\{+,-\}}(p)]+2\stackrel{\neg}{\lambda}^{S}_{\{+,-\}}(k)[\gamma^\alpha, \slashed{r}]\lambda^{S}_{\{+,-\}}(p)\stackrel{\neg}{\lambda}^{S}_{\{+,-\}}(k')\times[\gamma_\alpha, \slashed{r}](p'+k')\cdot(p+k)\lambda^{S}_{\{+,-\}}(p') \Big],
\end{eqnarray}
%\end{strip}
where $\mathcal{C}=\kappa^{2}/(256m^{2}r^{2})$.
\end{widetext}
Now we consider an elastic scattering in the center of mass frame, where $r^0=0$. We define the momentum for the participating fermions present in the interaction as $p^{\mu} = (E,0,p,0)$, $k^{\mu} = (E, psen\theta, pcos\theta, 0)$, $r^{\mu} = p^{\mu} - k^{\mu} = (0,-psen\theta, p - pcos\theta, 0)$, $p^{'\mu} = (E, 0, -p, 0)$ and $k^{'\mu} = (E, -psen\theta, -pcos\theta, 0)$. Thus, the amplitude for arbitrary momentum, after some manipulation, is given by
\begin{eqnarray}\label{3333}
 &&\mathcal{M}_{R} = \mathcal{C}\Big\{-256m^{6}+128m^{2}E^{2}(E^{2}+2p^{2}+2m^{2})\nonumber\\
 &&+128m^{2}p^{2}(p^{2}-2m^{2}cos\theta)
 +\frac{256m}{E+m}[(E+m)^{2}-p^{2}]\nonumber\\
 &&\times(2E^{2}-m^{2}+p^{2}(1+cos\theta))[Epcos(\theta/2)sen(\theta/2)\nonumber\\
 &&\times(sen\varphi - sen\theta cos\varphi - cos\theta sen\varphi)]-\frac{8(4E^{2}+p^{2})}{(E+m)^{2}}\nonumber\\
 &&\times[(E+m)^{2}-p^{2}]^{2}[4p^{2}(cos\theta -1)cos^{2}(\theta/2)sen^{2}(\theta/2)\nonumber\\
 &&(1+ cos\theta cos2\varphi - sen\theta sen2\varphi)]\Big\},\label{2111}
\end{eqnarray}
where ${r^{2}=2p^{2}(1-cos\theta)}$ and the subscript ${R}$ stands for `Relativistic'.

It is useful to perform the non-relativistic limit of the process at hand. As we shall see, it brings information about the low energy behavior of the Newtonian potential for mass dimension one fermions. This limit corresponds to take $(p\rightarrow 0, \theta\rightarrow 0)$ in the scattering amplitude. This procedure amounts out to $\mathcal{M}_{NR} = (\kappa^{2}/256m^{2}r^{2})\{128m^{6}\}$, where the subscript ${NR}$ means `Non-Relativistic'. Knowing that in the low energy context bosons and fermions are indistinguishable, we could expect that either the contributions are equal and come from both sectors (scalar and fermionic) or the contribution that comes from the fermionic sector is identically null (since the non-relativistic limit shall not distinguish anti-commuting properties). The second case is fulfilled here. The final amplitude reads 
$\mathcal{M}_{NR}=-\kappa^2/(256m^2{\bf r^2})128m^6 = -\kappa^2 m^4/(2{\bf r^2}) \Rightarrow -i\mathcal{M}_{NR} = i8\pi G_{grav}m^4/{\bf r^2}$. Now, by means of Eq. (\ref{v}), written in terms of spherical coordinates, we have
\begin{eqnarray}
V_{NR}(R) = -2\pi G_{grav}m^2\int \frac{\sin\theta d\theta d\varphi dr}{(2\pi)^3}e^{iRr\cos\theta},
\end{eqnarray}
leading to 
\begin{eqnarray}
 V_{NR}(R) = -\frac{G_{grav}m^2}{R},
\end{eqnarray}
reproducing the attractive Newtonian gravitational potential. This result may be faced as an additional support to the claim asserting mass dimension one spinors as dark matter candidates.

\section{Final Remarks}\label{FR}

Taking into account the peculiarities of the quantum field based upon eigenspinors of the charge conjugation operator, we studied here in some detail several relevant aspects of its coupling with gravity in the weak field regime. After evincing the specific interaction vertex we worked out the Ward-Takahashi identity. Going further, we evaluated the one-loop divergences of the graviton self-energy as well as the correct tadpole contribution, the former being canceled by the last one. Finally we perform the non-relativistic limit in order to show the appearance of the attractive Newtonian potential. We took special care in this limit, using in our calculations the polarized basis, already studied, and physically interpreting all important steps. 

The study composed by the first and third sections is quite important as it can in fact settle the fundamental aspects of semi-classical gravitational interaction for mass dimension one spinors. By its turn the investigation of the resulting non-relativistic gravitational potential as a consequence of the obtained vertex, since attractive, points to an adequate behavior under gravitational interaction from the perspective of a dark matter candidate. We hope that the results here discussed may serve also to push the investigations of mass dimension one spinors gravitational interaction even further. 

\begin{acknowledgments}
The authors thanks Prof. Jos\'e Abdalla Helayel-Neto for very stimulating and fruitful conversation. RJBR thanks to CNPq (Grant Number 155675/2018-4), RdCL and LCD thank to Coordena\c{c}\~ao de Aperfei\c{c}oamento de Pessoal de N\'ivel Superior - Brasil (CAPES) - Finance Code 001, and JMHdS thanks to CNPq (304629/2015-4; 303561/2018-1) for financial support.
\end{acknowledgments}

%%%%%%%%%%%%%%%%%%%%%%%%%%%%%%%%%%%%%%%%%%%%%%%%%%%%%%%%%%%%%%%%%%%%%%%%%%

\end{document}